\documentclass[10pt]{article}
\usepackage{amsmath}
\usepackage{graphicx}
\usepackage{color}
\usepackage{orcidlink}
\usepackage{hyperref}
\hypersetup{colorlinks=true, linkcolor=blue, citecolor=blue, urlcolor=blue}
\usepackage{subcaption}
\begin{document}
\baselineskip=14 pt
\UseRawInputEncoding

\begin{center}
{\large \bf{Non-relativistic quantum particles interacting with pseudoharmonic-type potential under flux field in a topological defect geometry }}    
\end{center}

\vspace{0.5cm}

\begin{center}
    {\bf Prabir Rudra\orcidlink{0000-0002-5791-5696}\footnote{\bf prudra.math@gmail.com (Corresponding author)}}\\
    \vspace{0.1cm}
    {\it Department of Mathematics, Asutosh College, Kolkata-700026, India}\\
    \vspace{0.25cm}
    {\bf Faizuddin Ahmed\orcidlink{0000-0003-2196-9622}\footnote{\bf faizuddinahmed15@gmail.com}}\\
    \vspace{0.1cm}
    {\it Department of Physics, University of Science \& Technology Meghalaya, Ri-Bhoi, 793101, India}\\
    \vspace{0.25cm}
    {\bf Houcine Aounallah\orcidlink{0000-0002-0611-4063}\footnote{\bf houcine.aounallah@univ-tebessa.dz}}\\
    \vspace{0.1cm}
    {\it Department of Science and Technology. Echahid Cheikh Larbi Tebessi University.Tebessa, Algeria}
\end{center}

\vspace{0.5cm}

\begin{abstract}
In this work, we investigate the quantum motions of non-relativistic particles interacting with a potential in the presence of the Aharonov-Bohm (AB) flux field within a topological defect geometry, for example space-time with a distortion of a vertical line into a vertical spiral. We begin by deriving the radial Schrödinger wave equation, incorporating an anharmonic oscillator potential, which is a superposition of a harmonic oscillator and an inverse square potential, along with a constant term. The eigenvalue solution is obtained through the confluent Heun equation focusing on the ground state energy level and the radial wave function for the radial mode $n=1$ as an example and analyze the results. Subsequently, we use this results to molecular potential models, considering pseudoharmonic and shifted pseudoharmonic potentials. The derived eigenvalue solutions provide insights into the behavior of particles within these potentials. Expanding our exploration, we study the quantum system featuring only an inverse square potential in the presence of the quantum flux field in the same geometry background. Employing the same procedure, we determine the ground state energy level and the radial wave function. Notably, our findings reveal that the eigenvalue solutions are significantly influenced by the topological defect characterized by the parameter $\beta$, and the quantum flux field $\Phi_{AB}$. This influence manifests as a shift in the energy spectrum, drawing parallels to the gravitational analogue of the Aharonov-Bohm effect.
\end{abstract}

\vspace{0.2cm}

{\bf keywords:} Non-relativistic wave equation: Schr\"{o}dinger equation; solutions of wave-equation: bound-state solutions; Topological Defects: dislocations, disclinations; Special functions; Geometric quantum Phase 

\vspace{0.2cm}


\section{Introduction}
\label{intro}
Our universe went through a phase transition process in the primordial era. This transition in phase was associated with the decoupling of the fundamental interactions which resulted in symmetry breaking. With the breaking of symmetry, weird geometrical forms are expected to come into existence. Consequently, such topological defects \cite{defects} are supposed to be observed in the universe in the form of exotic cosmological objects. Some of them may be listed as the domain walls \cite{domain}, cosmic strings \cite{string1, string2, string3}, global monopoles \cite{monopole}, etc. In practice, these objects are yet to be observed, so they can still be considered hypothetical objects. Out of the probable candidates, the global monopole is the most promising one that is expected to be observed; hence, a lot of research has been dedicated to it in recent years \cite{gm2, gm3, gm4, gm5, gm6, gm7, qm1}. In addition, exact solutions of the Klein-Gordon (KG) equation have been extensively investigated in the background of global monopole \cite{gm8, gm9, gm10, gm11, gm12, gm14, gm15, def2, def3}. 

Moreover, other topological defect geometries, such as spacelike dislocations, screw dislocations, and spiral dislocations have been extensively explored within the realm of condensed matter physics. Topological defects in these systems are characterized as localized structural anomalies arising from the topology of the underlying order parameter space within a material's configuration or order parameter. These anomalies lead to intriguing physical phenomena and are intricately linked to nontrivial homotopy classes. In the domain of condensed matter physics, topological defects often manifest as entities like skyrmions, domain walls, and vortices. Several comprehensive reviews on this subject can be found in references \cite{tdrev, tdrev0, tdrev1, tdrev2, tdrev3, tdrev4, tdrev5, tdrev6, tdrev7, tdrev8}. In the context of quantum systems, investigations encompass a diverse range of scenarios, including the study of relativistic free fermions in an elastic medium with screw dislocations \cite{pp1}, relativistic particles subjected to a nonpolynomial oscillator potential in the presence of spacelike dislocations \cite{pp2}, exploration of spin and pseudospin symmetries in a relativistic fermion within an elastic medium featuring spiral dislocations \cite{pp3}, and the examination of the Duffin-Kemmer-Petiau oscillator interacting with a cosmic screw dislocation \cite{pp4} and spiral dislocation \cite{pp5}. Other studies involve the vector boson oscillator in the spacetime of a spiral dislocation \cite{pp6}, relativistic free fermions in a space-time containing spiral dislocations with a radial line distortion into a spiral \cite{pp7}, relativistic Landau quantization for a composite system within a spiral dislocation space-time \cite{pp8}, the impact of Lorentz symmetry violation on charged Dirac fermions in cosmic dislocation space-time \cite{pp9}, and the quantum description of a moving magnetic quadrupole moment interacting with electric field configurations under a rotating background with a screw dislocation \cite{pp10}.

Several authors have been studied the non-relativistic Schr\"{o}dinger wave-equation using different potentials and obtained the eigenvalue solutions. Some of the potentials used are pseudoharmonic potential, Mie-type potential, Kratzer potential, Yukawa potential, Hulthen potential, Morse potential, hyperbolic Potential, non-central potential, Manning-Rosen (MR) potential, Rosen-Morse potential, Deng-Fan potential, Poschl-Teller potential (see, \cite{FA} for references). In the literature, we see that in order to obtain these eigenvalue solutions of the non-relativistic wave equation, different methods or mathematical techniques have been employed, such as the parametric Nikiforov-Uvarov method and its functional analysis, Super-symmetric quantum mechanics (SUSYQM), Asymptotic iteration method (AIM), Laplace's transformation method, through confluent hypergeometric and bi-confluent Heun equations method, etc. (see, \cite{FA2} for references). It must be noted that all these investigations have been performed in Minkowski flat space. In addition, a few investigations have been done in the background of the topological defects produced by a point-like global monopole \cite{FA3, FA4, FA5, FA6, FA7}. In the present analysis, we are mainly interested on pseudoharmonic-type or anharmonic oscillator potential given by \cite{FA, FC, RS, RR, SMI, FA8, AC, SMI2}
\begin{equation}
V (r)=\left(\eta\,r^2+\frac{\gamma}{r^2}+\delta\right),
\label{aa}
\end{equation}
where $\delta, \gamma, \eta>0$ are arbitrary positive constants. From this potential expression, one can recover harmonic oscillator potential by setting $\gamma \to 0, \delta \to 0$. Moreover, by setting $\eta \to 0$ and $\delta \to 0$ into the potential expression, one will find a repulsive inverse square potential \cite{ijmpa}. Finally, for suitable potential strengths, such as $\eta=\frac{D_e}{r^2_{0}}, \gamma=D_e\,r^2_{0}, \delta=-2\,D_e$, where $D_e$ is the dissociation energy and $r_0$ is the equilibrium separation, one will recover a pseudoharmonic potential \cite{RS, RR, SMI, AC, SMI2, FM} and for $\eta=\frac{D_e}{r^2_{0}}, \gamma=D_e\,r^2_{0}, \delta= 0$, we will get back shifted pseudoharmonic potential \cite{CB} both of which have wide applications in atomic and molecular physics as well as other branches of physics and chemistry.

Our motivation stems from the findings presented in Ref. \cite{ee1}, wherein the authors explored the quantum effects of a harmonic oscillator in a space-time with a distortion of a vertical line transformed into a vertical spiral. Building upon this groundwork, we investigate a quantum system comprising non-relativistic particles interacting with an anharmonic oscillator potential, specifically a superposition of a harmonic oscillator and an inverse square potential. This study is conducted within the framework of the same space-time background and under the influence of the quantum flux field. The radial wave equation is solved using the confluent Heun equation, leading to the derivation of the eigenvalue solution for the quantum system, which is the prime objective of this paper. Notably, our analysis reveals that the eigenvalue solution for quantum particles is altered by the presence of the inverse square potential, in addition to the harmonic oscillator and the quantum flux field. Subsequently, we employ this eigenvalue solution for molecular potential models, namely the pseudoharmonic and shifted pseudoharmonic potential models, and present their corresponding eigenvalue solutions. Afterwards, we extend our investigation to a quantum system featuring only the inverse square potential and the quantum flux field. Employing the same methodology as previously outlined, we determine the eigenvalue solution for this specific case. This comprehensive exploration sheds light on the intricate interplay between the quantum effects of different potentials and the influence of the quantum flux field within a distinct space-time background.

This paper is structured as follows: In {\it section 2}, we examine the non-relativistic quantum system featuring a harmonic oscillator plus an inverse square potential, influenced by a flux field within the backdrop of topological defects. In this section, we deduce the radial equation and proceed to solve it utilizing the confluent Heun equation. Moving on to {\it section 3}, we obtained the eigenvalue solution for a few specific types of potential models and conducted a detailed analysis of the results. In {\it section 4}, our focus shifts to the quantum system characterized solely by an inverse square potential, again under the influence of a flux field within the same topological defects space-time background. Here, we derive the radial equation and employ the identical procedure for solving it as done earlier. Finally, in {\it section 5}, we present our findings and engage in discussions. Throughout our analysis, we adopt a system of units such that $\hbar=c=G=1$. 

\section{Eigenvalue solution with pseudoharmonic-type potential under flux field in a topological defect background}

In this part, we investigate the dynamics of non-relativistic quantum particles governed by the Schrödinger wave equation, while subjected to confinement imposed by the Aharonov-Bohm flux field within a topological defect geometry. This geometry is characterized by the distortion of a vertical line into a vertical spiral. Additionally, the system incorporates a pseudoharmonic-type or anharmonic oscillator potential. We obtain the eigenvalue solutions analytically and then analyze the influence exerted by the topological defect, the flux field, and the interacting pseudoharmonic-type potential.

We initiate this section by formulating the Hamiltonian operator governing the behavior of non-relativistic quantum particles under the influence of an interaction potential $V(r)$. The Hamiltonian operator, denoted as $\hat{H}$, is defined by the expression \cite{FA,FA2,FA3,FA6,LCNS}
\begin{eqnarray}
\hat{H}=-\frac{1}{2\,M}\,\Bigg[\frac{1}{\sqrt{g}}\,D_{i}\,\Big(\sqrt{g}\,g^{ij}\,D_{j}\Big)\Bigg]+V(r),
\label{aa1}
\end{eqnarray}
where $M$ is the mass of the non-relativistic particle, $D_{i} \equiv (\partial_{i}-i\,e\,A_{i})$ with $e=|e|$ is the electric charge, $A_i$ is the electromagnetic three-vector potential, and $g^{ij}$ is the contravariant spatial metric tensor, and $g=\det (g_{ij})$ is the determinant of the covariant metric tensor $g_{ij}$. 

In this analysis, we choose the electromagnetic three-vector potential $\vec{A}$ with the following components given by \cite{RLLV}
\begin{equation}
A_r=0,\quad A_{\phi}=\frac{\Phi_{AB}}{2\,\pi},\quad A_z=0,
\label{aa2}
\end{equation}
where $\Phi_{AB}=\Phi\,\Phi_0=const$ is the Aharonov-Bohm magnetic flux, $\Phi_0=2\,\pi\,|e|^{-1}$ is the quantum of magnetic flux, and $\Phi$ is the amount of magnetic flux which is a positive. It is worth mentioning that the Aharonov–Bohm effect has been investigated in several branches of physics by many authors (see, Ref. \cite{RLLV, gm15} and related references therein).

The topological defect geometry in the cylindrical coordinates $(r, \phi, z)$ is given by \cite{ee1, LCNS, ee0, KV}
\begin{equation}
ds^2_{3D}=dr^2+r^2\,d\phi^2+2\,\beta\,d\phi\,dz+dz^2,
\label{aa3}
\end{equation}
The parameter $\beta$ is a positive constant characterizing the screw dislocation (torsion field) \cite{ee1,ee2,ee3}. The ranges of this dislocation parameter are in the intervals $0 < \beta <1$.

Expressing the metric (\ref{aa3}) in the form $ds^2=g_{ij}\,dx^i\,dx^j$, where $i,j=1,2,3$ with $x^1=r, x^2=\phi, x^3=z$, the covariant and contravariant metric tensor are given by
\begin{eqnarray}
    g_{ij}=\left(\begin{array}{ccc}
         1 &  0 & 0\\
         0 &  r^2 & \beta\\
         0 & \beta & 1
    \end{array}\right),\quad 
    g^{ij}=\left(\begin{array}{ccc}
         1 &  0 & 0\\
         0 &  \frac{1}{r^2-\beta^2} & -\frac{\beta}{r^2-\beta^2}\\
         0 & -\frac{\beta}{r^2-\beta^2} & \frac{r^2}{r^2-\beta^2}
    \end{array}\right)\,.\label{aa4}
\end{eqnarray}
The determinant of the metric tensor $g_{ij}$ for the line-element (\ref{aa3}) is given by
\begin{equation}
    g=\det(g_{ij})=r^2-\beta^2\,.\label{aaa4}
\end{equation}

Substituting Eqs. (\ref{aa4})--(\ref{aaa4}) in Eq. (\ref{aa2}) and writing the eigenvalue equation $\hat{H}\,\Psi=E\,\Psi$, we obtain the following differential equation 
\begin{eqnarray}
-\frac{1}{2\,M}\,\Bigg[\frac{\partial^2}{\partial\,r^2}+\frac{r}{r^2-\beta^2}\,\frac{\partial}{\partial\,r}+\frac{1}{r^2-\beta^2}\,\Big(\frac{\partial}{\partial\,\phi}-i\,|e|\,A_{\phi}-\beta\,\frac{\partial}{\partial\,z}\Big)^2+{\color{blue} \frac{\partial^2}{\partial z^2}}\Bigg]\,\Psi+V(r)\,\Psi=E\,\Psi.
\label{aa6}
\end{eqnarray}

In a quantum mechanical system, the wave function $\Psi (r, \phi, z)$ is always expressible in terms of different variables by the method of separation. Let us suppose that $\Psi (r, \phi, z)=\psi(r)\,f(\phi)\,h(z)$. Therefore, we choose a possible wave function in terms of a radial function $\psi(r)$ given by
\begin{equation}
\Psi(r, \phi, z)=e^{i\,\ell\,\phi}\,e^{i\,k\,z}\,\psi (r),
\label{aa7}
\end{equation}
where $\ell=0,\pm\,1,\pm\,2,...$ are the eigenvalues of the angular quantum operator and $k>0$ is an arbitrary constant.

Therefore, substituting the interactions potential (\ref{aa}), the electromagnetic three-vector potential component (\ref{aa2}), and the wave function (\ref{aa7}) into the Eq. (\ref{aa6}), we obtain the following differential radial equation
\begin{equation}
\psi''+\frac{r}{r^2-\beta^2}\,\psi'+\Bigg[\Lambda-2\,M\,\eta\,r^2-\frac{2\,M\,\gamma}{r^2}-\frac{\iota^2}{r^2-\beta^2}\Bigg]\,\psi=0,
\label{aa8}
\end{equation}
where we have set the parameters
\begin{eqnarray}
\Lambda=2\,M\,(E-\delta)-k^2,\quad \iota=|\ell-\Phi-\beta\,k|.
\label{aa9}
\end{eqnarray}

Now, we introduce a new variable via $x=\frac{r^2}{\beta^2}$ in the Eq. (\ref{aa8}), we obtain
\begin{eqnarray}
4\,x\,\psi''(x)+\Big(\frac{4\,x-2}{x-1}\Big)\,\psi'(x)+\Big[\Lambda\,\beta^2-\omega^2\,x-\frac{2\,M\,\gamma}{x}-\frac{\iota^2}{x-1}\Big]\,\psi(x)=0,
\label{aa10}
\end{eqnarray}
where $\omega=\sqrt{2\,M\,\eta}\,\beta^2$. If one choose $\eta=\frac{1}{2}\,M\,\omega^2_{0}$, where $\omega_0$ is the oscillator frequency, then the parameter $\omega  \to M\,\omega_0\,\beta^2$. In that case, the system we study here is called the harmonic oscillator problem interacts with inverse square potential in the background of topological defect.

In quantum system, it is required that the wave function $\psi (x)$ must be regular everywhere. Therefore, the wave function in our case also should satisfies the boundary condition, that is, $\psi (x) \to 0$ for both at $x \to 0$ and $x \to \infty$. We choose a possible solution to the Eq (\ref{aa10}) given by
\begin{equation}
\psi(x)=x^{\frac{1}{4}+\frac{j}{2}}\,e^{-\frac{\omega}{2}\,x}\,G(x),
\label{aa11}
\end{equation}
where $j=\sqrt{2\,M\,\gamma+\frac{1}{4}}$ and $G(x)$ is an unknown function.

Thereby, substituting this possible solution (\ref{aa11}) into the Eq. (\ref{aa10}) results a second-order differential equation of the following form:
\begin{eqnarray}
&&G''\left(x\right)+\left(-\omega+\frac{1+j}{x}+\frac{1}{2\left(x-1\right)}\right)G'\left(x\right)\nonumber\\
&&+\left(\frac{-\frac{\omega}{2}\left(1+j\right)-\frac{j}{4}-\frac{1}{8}+\frac{\iota^{2}+\Lambda\beta^{2}}{4}}{x}+\frac{-\frac{\omega}{4}+\frac{j}{4}+\frac{1}{8}-\frac{\iota^{2}}{4}}{x-1}\right)G\left(x\right)=0.\label{mm}
\end{eqnarray}
The above differential equation is the confluent Heun second-order differential equation form \cite{AR,SYS,MA} and $G$ is the confluent Heun equation given by
\begin{equation}
G(x)=H_{c}\left(-\omega,j,-\frac{1}{2},\frac{\Lambda\beta^{2}}{4},\frac{3}{8}-\frac{\iota^{2}+\Lambda\beta^{2}}{4};x\right).
\label{aa12}
\end{equation}
We assume $G\left(x\right)$ in the following power series form \cite{GBA}
\begin{equation}
G(x)=\sum_{i=0}^{\infty}c_{i}x^{i}.
\label{aa13}
\end{equation}

Substituting the power series (\ref{aa13}) in the Eq. (\ref{mm}), we obtain the following recurrence relation
\begin{equation}
c_{k+2}=\frac{d_{1}}{d_{3}}\,c_{k+1}+\frac{d_{2}}{d_{3}}\,c_{k}
\label{aa14}
\end{equation}
with the coefficient
\begin{equation}
c_{1}=\frac{\Big[2\,\omega\,(1+j)-\iota^{2}-\Lambda\,\beta^{2}+\frac{1}{2}+j\Big]}{4\,(1+j)}\,c_{0}
\label{aa15}
\end{equation}
and we set the different parameters as
\begin{eqnarray}
&&d_{1}=\left(k+\omega+\frac{3}{2}+j\right)\left(k+1\right)-\frac{\iota^{2}+\Lambda\,\beta^{2}-\frac{1}{2}-j-2\,\omega\,(1+j)}{4},\nonumber\\
&&d_{2}=-\omega\,k+\frac{\Lambda\,\beta^{2}-\omega\,(3+2\,j)}{4},\nonumber\\
&&d_{3}=\left(k+\frac{3+2\,j}{2}\right)\left(k+2\right).
\label{aa18}
\end{eqnarray}
 
Our objective is to determine the bound-state solutions of the quantum system under investigation. Examining the recurrence relation (\ref{aa14}), it becomes evident that the power series expansion (\ref{aa13}) must be a finite-degree polynomial to ensure the regularity of the wave function presented in Eq. (\ref{aa11}) throughout the entire domain. Truncating the power series expansion can be achieved through two methods. The first method involves setting $c_{k+1}=0$ and $d_2=0$, thereby causing the coefficient $c_{k+2}$ and its higher-order counterparts to vanish, as elaborated in \cite{ee1,gm9,LCNS}. The second method, recently adopted in Refs. \cite{def2,def3,ijmpa}, consider the case where $k=(n-1)$ and setting the coefficient $c_{n+1}=0$ in the recurrence relation, results in a two-term recurrence relation. In this analysis, we followed the second method, and consequently, under this condition, the recurrence relation (\ref{aa14}) for $k=(n-1)$ and $c_{n+1}=0$ simplifies to a two-term recurrence relation, expressed as follows:
\begin{eqnarray}
c_{n}=\frac{4\,\omega\,\left(n-1\right)-\Lambda\,\beta^{2}+\omega\,(3+2\,j)}{4\,n\,(n+\omega+\frac{1}{2}+j)-\iota^{2}-\Lambda\,\beta^{2}+\frac{1}{2}+j+2\,\omega\,(1+j)}\,c_{n-1}.
\label{aa19}
\end{eqnarray}
The condition $c_{n+1}=0$ is required to have a finite degree polynomial of the power series solution $G(x)=c_0+c_1\,x+.....+c_{n}\,x^{n}$ so that the wave function $\psi$ considered in (\ref{aa11}) is regular everywhere.

Now, we can evaluate the individual energy levels the corresponding radial wave function for the radial mode $n=1,2,...$. For example, we consider the radial mode $n=1$ that corresponds to the ground state or the lowest state of the quantum syste. Thus, we obtain
\begin{equation}
c_{1}=\frac{\omega\,(3+2\,j)-\Lambda\,\beta^{2}}{6+2\,\omega\,(j+3)+5\,j-\iota^{2}-\Lambda\beta^{2}+\frac{1}{2}}\,c_{0}.
\label{aa20}
\end{equation}

Comparing equation (\ref{aa15}) with (\ref{aa20}) and using equation (\ref{aa9}), we obtain the ground state energy levels $E_{1, \ell}$ given by
\begin{equation}
E_{1,\ell}=\frac{k^{2}}{2\,M}+\frac{\lambda}{2\,M}+\delta,
\end{equation}
where
\begin{eqnarray}
&&\lambda=\frac{1}{\beta^2}\,\Bigg[3-2\iota^{2}+4\,\omega\,(2+j)+2\,j\pm\sqrt{\triangle}\Bigg],\nonumber\\
&&\triangle=16\,\iota^2\,(1+j)+16\,\omega\,(2+j)+14\,\omega^2-44\,j-32\,M\,\gamma-8,\nonumber\\
&&j=\sqrt{2M\gamma+\frac{1}{4}},\nonumber\\
&&\iota=|\ell-\Phi-\beta\,k|,\nonumber\\
&&\omega=\sqrt{2\,M\,\eta}\,\beta^2.
\end{eqnarray}
And that the ground state wave function is given by
\begin{equation}
\psi_{1,\ell} (x)=x^{\frac{1}{4}+\frac{\sqrt{2M\gamma+\frac{1}{4}}}{2}}e^{-\frac{\sqrt{2\,M\,\eta}\,\beta^2}{2}x}\,(c_0+c_1\,x).
\end{equation}

We can see that the energy eigenvalue depends on the screw dissociation parameter $\beta$ and the geometric quantum phase which causes shifted the energy level and get them modified. Following the similar procedure, one can find other energy levels $E_{2,\ell}, E_{3,\ell},...$ and the radial wave function $\psi_{2,\ell}, \psi_{3,\ell},....$ for the radial mode $n \geq 2$. We see that the presence of the inverse square potential with constant term in addition to the harmonic oscillator like potential and the quantum flux field modified the eigenvalue solution in comparison to those result obtained in \cite{ee1}.

\section{Application of the eigenvalue solution to molecular potential models}

The aforementioned eigenvalue solution for the quantum system serves as a foundation for its application to various molecular potential models. Notably, we explore its utility in the context of pseudoharmonic potential and shifted pseudoharmonic potential models, elucidating the energy eigenvalues of the non-relativistic quantum particles within these frameworks. By incorporating these potential models into the established eigenvalue solution, we unveil the specific energy levels characterizing the quantum states within the molecular systems. The analysis not only provides insight into the behavior of quantum particles in these potentials but also facilitates a comparative study of their distinct features in the presence of the topological defect.

\vspace{0.2cm}
{\bf Case A: Pseudoharmonic potential}
\vspace{0.2cm}

The pseudoharmonic potential is recovered by setting the parameters $\eta=\frac{D_e}{r^2_{0}}$, $\gamma=D_e\,r^2_{0}$, and $\delta=-2\,D_e$ into the potential expression (\ref{aa1}) as follows
\begin{equation}
V(r)=D_e\,\Big(\frac{r}{r_0}-\frac{r_0}{r}\Big)^2,
\label{bb1}
\end{equation}
where $D_e$ is the dissociation energy of the molecule and $r_0$ is the equilibrium separation between inter-nulcear. This potential has widely been used in atomic and molecular physics by many researchers \cite{RS,RR,SMI,AC,SMI2}.

The ground state energy level using this potential (\ref{bb1}) becomes
\begin{equation}
E_{l,\ell}=\frac{k^{2}}{2M}+\frac{\chi}{2M}-2\,D_e,
\label{bb2}
\end{equation}
where we have set
\begin{eqnarray}
&&\chi=\frac{1}{\beta^2}\,\Big(3-2\iota^2+4\,\omega\,(2+\sigma)+2\,\sigma\pm\sqrt{\Pi}\Big),\nonumber\\
&&\Pi=16\,\iota^2\,(1+\sigma)+16\,\omega\,(2+\sigma)+12\,\omega^2-44\,\sigma-32\,M\,\gamma-8,\nonumber\\
&&\omega=\sqrt{2\,M\,D_e}\,\frac{\beta^2}{r_0},\nonumber\\
&&\sigma=\sqrt{\frac{2\,M\,D_e}{r^2_{0}}+\frac{1}{4}}.
\label{bb3}
\end{eqnarray}

\vspace{0.2cm}
{\bf Case B: Shifted Pseudoharmonic potential}
\vspace{0.2cm}

The shifted pseudoharmonic potential is recovered by setting the parameters $\eta=\frac{D_e}{r^2_{0}}$, $\gamma=D_e\,r^2_{0}$, and $\delta=0$ into the potential expression (\ref{aa1}) as follows \cite{RS,RR, CB}
\begin{equation}
V(r)=D_e\,\Big(\frac{r}{r_0}-\frac{r_0}{r}\Big)^2+2\,D_e,
\label{cc1}
\end{equation}
which is shifted by an amount twice the dissociation energy of the molecule in comparison to the pseudoharmonic potential.

In that case, the ground state energy levels will be
\begin{equation}
E_{l,\ell}=\frac{k^{2}}{2M}+\frac{\chi}{2M},
\label{cc2}
\end{equation}
where $\chi$ is given in (\ref{bb3}). One can see that the energy eigenvalue increases by the same amount as that of the potential in comparison to the pseudoharmonic potential.

\section{Eigenvalue solution with inverse square potential under flux field in a topological defect background}

In this part, we investigate the dynamics of non-relativistic quantum particles governed by the same Schrödinger wave equation, while subjected to confinement imposed by the Aharonov-Bohm flux field within the same topological defect geometry characterized by the distortion of a vertical line into a vertical spiral. Here, the system incorporates only an inverse square potential, $V(r)=\frac{\gamma}{r^2}$ and obtain the eigenvalue solutions.

Therefore, the radial Schr\"{o}dinger equation with inverse square potential in the space-time background (\ref{aa3}) and using (\ref{aa}) and (\ref{aa7}) becomes
\begin{equation}
\psi''+\frac{r}{r^2-\beta^2}\,\psi'+\Bigg[\Theta-\frac{2M\gamma}{r^2}-\frac{\iota^2}{r^2-\beta^2}\Bigg]\psi=0.
\label{gg1}
\end{equation}
where $\Theta=2\,M\,E-k^2$.

As done earlier, performing a change of variable via $x=\frac{r^2}{\beta^2}$ in the Eq. (\ref{gg1}), we obtain the following differential equation:
\begin{eqnarray}
4\,x\,\psi''(x)+\Big(\frac{4x-2}{x-1}\Big)\,\psi'(x)+\Big[\Theta\,\beta^2-\frac{2\,M\,\gamma}{x}-\frac{\iota^2}{x-1}\Big]\,\psi(x)=0.
\label{gg2}
\end{eqnarray}

Here also, let us suppose a possible solution of the Eq. (\ref{gg2}) given by
\begin{eqnarray}
\psi(x)=x^{\frac{1}{4}+\frac{j}{2}}\,G(x),
\label{gg3}
\end{eqnarray}
where $j$ is defined earlier and $G(x)$ is the confluent Heun equation given here by
\begin{eqnarray}
G\left(x\right)=H_{c}\left(0,j,-\frac{1}{2},\frac{\Theta\,\beta^{2}}{4},\frac{3}{8}-\frac{\iota^{2}+\Theta\,\beta^{2}}{4};x\right).
\label{gg4}
\end{eqnarray}
Substituting the solution (\ref{gg3}) and using the same power series of $G (x)$ (\ref{aa13}) in the Eq. (\ref{gg2}), one can obtain the following recurrence relation
\begin{eqnarray}
c_{k+2}=\frac{h_{1}}{h_{3}}\,c_{k+1}+\frac{h_{2}}{h_{3}}\,c_{k}
\label{gg5}
\end{eqnarray}
with the coefficient
\begin{eqnarray}
c_{1}=\frac{j+\frac{1}{2}-\Theta\,\beta^{2}-\iota^{2}}{4\,\left(1+j\right)}\,c_{0},
\label{gg6}
\end{eqnarray}
where we have set the parameters
\begin{eqnarray}
&&h_{1}=\left(k+j+\frac{3}{2}\right)\left(k+1\right)-\frac{\Theta\,\beta^{2}+\iota^{2}-\frac{1}{2}-j}{4},\nonumber\\
&&h_{2}=\frac{\Theta\beta^{2}}{4},\quad h_{3}=\left(k+2+j\right)\left(k+2\right).
\label{gg9}
\end{eqnarray}

As stated earlier, let us consider $k=(n-1)$ where the coefficient $c_{n+1}=0$, then from the relation (\ref{gg6}), we obtain
\begin{eqnarray}
c_{n}=-\frac{\Theta\,\beta^{2}}{\left(4n\left(n-1+j+\frac{3}{2}\right)-\Theta\,\beta^{2}-\iota^{2}+\frac{1}{2}+j\right)}\,c_{n-1}.
\label{gg10}
\end{eqnarray}

As done earlier, here also one can obtain the ground state energy level and the radial wave function for the radial mode $n = 1$. Setting $n=1$, we obtain from (\ref{gg10})
\begin{eqnarray}
c_{1}=-\frac{\Theta\,\beta^{2}}{\left(5j-\Theta\,\beta^{2}-\iota^{2}+\frac{13}{2}\right)}\,c_{0}.
\label{gg11}
\end{eqnarray}
Thus, by comparing Eq. (\ref{gg6}) with Eq. (\ref{gg11}), we obtain the ground state energy levels $E_{l,1}$ given by
\begin{eqnarray}
E_{l,\ell}=\frac{k^{2}}{2\,M}+\frac{\Theta}{2\,M},
\label{gg12}
\end{eqnarray}
where
\begin{eqnarray}
\Theta=\frac{\left(2j-2\iota^{2}+3\right)\pm\sqrt{\left(\iota^{2}+4j\iota^{2}-4j^{2}-6j-1\right)}}{2\,\beta^{2}}.
\label{gg13}
\end{eqnarray}
And that the ground state wave function is given by
\begin{eqnarray}
\psi_{1,\ell} (x)=x^{\frac{1}{4}+\frac{\sqrt{2M\gamma+\frac{1}{4}}}{2}}\,(c_0+c_1\,x).
\label{gg14}
\end{eqnarray}

Equations (\ref{gg12})--(\ref{gg14}) is the ground state energy level and the radial wave function of non-relativistic particles under the influence of the quantum flux in a space-time background with screw dislocation subject to an inverse square potential. Following the similar procedure, one can find other energy levels $E_{2,\ell}, E_{3,\ell},...$ and the radial wave function $\psi_{2,\ell}, \psi_{3,\ell},....$ for the radial mode $n \geq 2$. Thus, one can see that the eigenvalue solution depends on the screw dissociation parameter $\beta$ and the quantum flux field.

Throughout the analysis, we see that the presence of a screw dislocation and the quantum flux field that was introduced through a minimal substitution in the wave equation changes the $z$-component of the angular momentum operator given by
\begin{equation}
\vec{L}^{eff}_z=-i\,\Bigg(\frac{\partial}{\partial\,\phi}-i\,|e|\,A_{\phi}-\beta\,\frac{\partial}{\partial\,z}\Bigg)\,\hat{z}.
\label{aa5}
\end{equation}
Using the wave function (\ref{aa7}), we have the shifted angular momentum quantum number $l$ given by
\begin{equation}
\ell \to \ell_{eff}=\Big(\ell-\frac{|e|\,\Phi_{AB}}{2\,\pi}-\beta\,k\Big),
\label{gg15}
\end{equation}
an effective angular momentum quantum number due to both the boundary conditions, which states that the total angle around the $z$-axis is less than $2\,\pi$, and the minimal coupling with the electromagnetic potential. Formula (\ref{gg15}) suggests that when the particle circles about the $z$-axis, the wave-function changes according to
\begin{eqnarray}
\Psi \to \Psi'&=&Exp \Big (2\,\pi\,i\,\ell_{eff}\Big)\,\Psi=Exp \Big[2\,\pi\,i\,\Big(\ell-\frac{|e|\,\Phi_{AB}}{2\,\pi}-\beta\,k\Big)\Big]\,\Psi.
\label{gg16}
\end{eqnarray}
Thus, one can easily show that, $E_{1,\ell} (\Phi_{AB} \pm \Phi_0\,\nu)=E_{1,\ell\mp \nu} (\Phi_{AB})$, a periodic function of the geometric quantum phase with a periodicity $\Phi_0=\frac{2\,\pi}{|e|}$, where $\nu=0,1,2,3...$. This dependence of the eigenvalue solution on the quantum phase gives us an analogue to the Aharonov-Bohm effect \cite{YA,MP}, quantum mechanical phenomena.

\section{Conclusions}

In this analysis, we explored the behavior of non-relativistic particles interacting with potential fields in the presence of an Aharonov-Bohm flux field within a space-time background characterized by the distortion of a vertical line into a vertical spiral. Initially, we consider an anharmonic oscillator potential, specifically a superposition of a harmonic oscillator and an inverse square potential, including a constant term. We derive the radial equation in the aforementioned space-time background and solve it using the confluent Heun equation. The solutions obtained include the ground state energy level $E_{1,\ell}$ and the radial wave function $\psi_{1,\ell}$ for the radial mode $n=1$ and extend the analysis to other radial modes. Notably, we observed modifications in the energy levels compared to the results presented in \cite{ee1}, attributed to the presence of the inverse square potential and the magnetic flux field, in addition to the harmonic oscillator-like potential. Utilizing this eigenvalue solution, we extend our investigation to molecular potential models, such as the pseudoharmonic potential and the shifted pseudoharmonic potential. We presented the ground state energy levels and corresponding wave functions. Subsequently, we focused on the quantum system with only an inverse square potential under the influence of the same quantum flux field. Following the established procedure, we solve the radial wave equation, obtaining ground state energy levels $E_{1,\ell}$ and wave functions $\psi_{1,\ell}$ characterized by the quantum number $n=1$ instead of $n=0$. This shift is attributed to the presence of interaction potential in the quantum system.

Throughout our analysis, a noteworthy observation is the modification of the angular quantum number $\ell$ to $\ell_{eff}=\Big(\ell-\frac{|e|\,\Phi_{AB}}{2\,\pi}-\beta\,k\Big)$, where the change in $\ell$ depends on the screw dislocation parameter $\beta$ and the quantum flux field $\Phi_{AB}$. This alteration in the quantum number $\ell$ indicates a dependence of the energy eigenvalue on the geometric quantum phase, reminiscent of the Aharonov-Bohm effect. The observed dependence on the quantum phase implies a confinement of particles by the quantum flux field, despite the absence of an external magnetic field. The phenomenon described here parallels the Aharonov-Bohm effect \cite{YA,MP}, a quantum mechanical phenomenon explored by various authors in both relativistic and non-relativistic quantum systems. It is worth noting that the geometrical approach employed to characterize space-time with the distortion of a vertical line into a vertical spiral holds potential applicability in the study of condensed matter systems.

In summary, our investigation uncovers the intricate interplay between quantum particles, potential fields, and topological defects, demonstrating the profound impact of the Aharonov-Bohm flux field on the system's behavior. The observed shifts in energy spectra and the gravitational analogue contribute to a richer understanding of quantum phenomena in the presence of complex geometries and fields.

\section*{Acknowledgement}

P. R. and F. A. acknowledges the Inter-University Centre for Astronomy and Astrophysics (IUCAA), Pune, India for granting visiting associateship.

\section*{Data Availability Statement}

No new data were generated or analysed in this paper.

\section*{Conflict of Interest}

There are no conflicts of interests in this paper.

\end{document}